\documentclass[12pt]{iopart}
\usepackage{iopams}  
\usepackage{natbib}
\usepackage{graphicx}

\begin{document}

\title[Millisecond Pulsar Searches]{Searching for Millisecond
Pulsars: Surveys, Techniques and Prospects}

\author{K. Stovall$^{1,2}$, D.R. Lorimer$^{3,4,5}$, R.S. Lynch$^6$}

\address{$^1$Center for Advanced Radio Astronomy, University of Texas at Brownsville, 80 Fort Brown, Brownsville, TX 78520, USA}
\address{$^2$Department of Physics and Astronomy, University of Texas at San Antonio, One UTSA Circle, San Antonio, TX 78249, USA}
\address{$^3$Department of Physics, West Virginia University, White Hall, Morgantown, WV 26506, USA}
\address{$^4$National Radio Astronomy Observatory, PO Box 2, Green Bank, WV 24944, USA}
\address{$^5$Astrophysics, University of Oxford, Denys Wilkinson Building, Keble Road, Oxford OX1 3RH, UK}
\address{$^6$Department of Physics, McGill University, 3600 University Street, Montreal, QC H3A 2T8, Canada}
\ead{stovall.kevin@gmail.com}

\begin{abstract}
Searches for millisecond pulsars (which we here loosely define as
those with periods $<$~20~ms) in the Galactic field have undergone a
renaissance in the past five years. New or recently refurbished radio
telescopes utilizing cooled receivers and state-of-the art digital
data acquisition systems are carrying out surveys of the entire sky at
a variety of radio frequencies.  Targeted searches for millisecond
pulsars in point sources identified by the {\it Fermi} Gamma-ray
Space Telescope have proved phenomenally successful, with over 50
discoveries in the past five years. The current sample of millisecond
pulsars now numbers almost 200 and, for the first time in 25 years,
now outnumbers their counterparts in Galactic globular clusters. While
many of these searches are motivated to find pulsars which form part
of pulsar timing arrays, a wide variety of interesting systems are now
being found.  Following a brief overview of the millisecond pulsar
phenomenon, we describe these searches and present some of the
highlights of the new discoveries in the past decade. We conclude with
predictions and prospects for ongoing and future surveys.
\end{abstract}

\section{Introduction}\label{sec:intro}

Since their discovery 45 years ago \citep{hbp+68}, just over 2000
pulsars have been found, enabling some of the most
fascinating astronomical discoveries over that same time span.  The
majority of these are radio pulsars that have been found in large-area
surveys, though a significant fraction, especially of millisecond
pulsars (MSPs), were found in targeted searches.  Large-area surveys
entail a methodic search of large, generally contiguous, regions of
the sky for pulsars; while targeted searches involve the search of
a known object, such as a gamma-ray source or a supernova remnant,
for the detection of a pulsar.  Because we are still
only sampling a small fraction of the underlying population, almost
all surveys result in some new and often unexpected discovery, many of
which have an impact beyond the astrophysical study of neutron stars.  Some
highlights from the past few years are: the double pulsar J0737-3039
\citep{lbk+04}, which has provided some of the best tests of
strong-field general relativity \citep{ksm+06}, as well pulsar-white
dwarf systems that place stringent limits on tensor-vector-scalar
theories of gravity \citep{bbv08,lwj+09,fwe+12,afw+13}; PSR
J1614$-$2230, a 2 solar mass neutron star that has provided the
best constraints yet on the equation of state of ultra-dense matter
\citep{dpr+10}; and the unexpected discovery of a population of
gamma-ray pulsars and MSPs \citep{aaa+09}.  As discussed by
several other authors in this focus issue, e.g. the article by
R.~N.~Manchester, one of the most exciting prospects is the direct
detection of gravitational waves using a
pulsar timing array (PTA) of precision MSPs, which will help to usher
in a new era of gravitational wave astronomy.

\begin{figure}[h!]
\includegraphics[width=\textwidth]{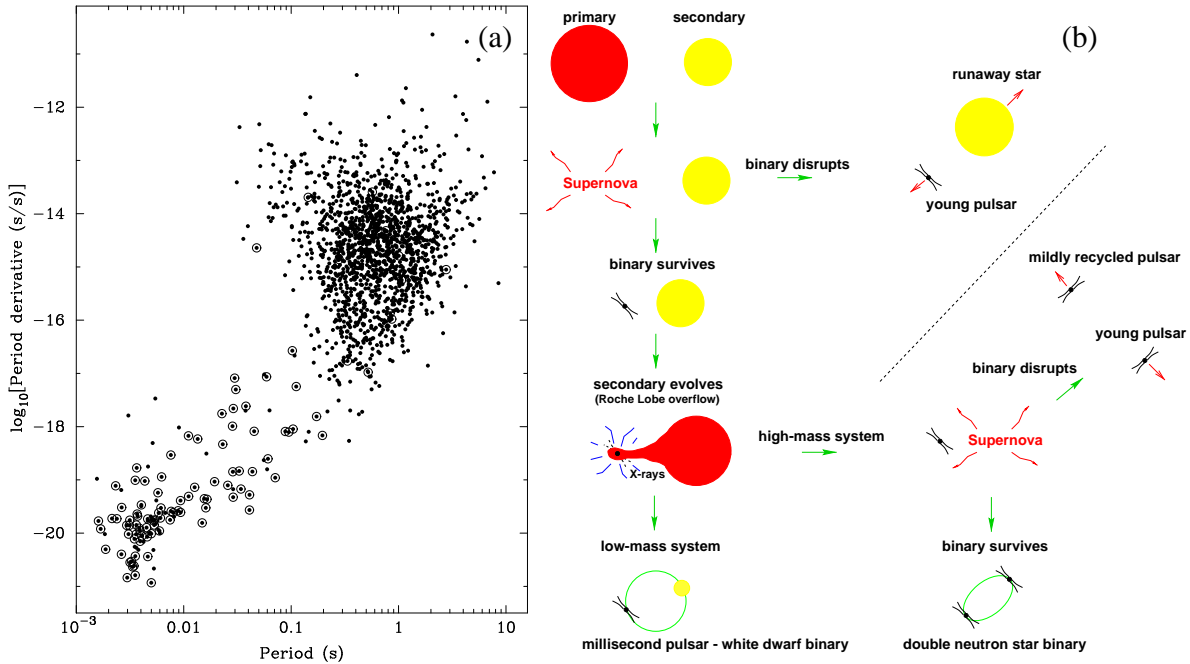}
\caption{\em \label{fig:evolution} (a): $P-\dot{P}$ diagram
showing the current sample of radio pulsars. Binary pulsars are
highlighted by open circles. (b): Binary evolution scenarios involving
pulsars (see text)~\citep{lor08}.
}
\end{figure}

Searching for MSPs is fraught with unique challenges that do not
affect searches for long-period pulsars to the same degree.  The
sampling rate necessary to detect MSPs is high, as is the required
spectral resolution.  The latter is needed to overcome the deleterious
effects of dispersive smearing by free electrons in the interstellar
medium. While long-period pulsars can be detected with of order 100 channels
sampled every few milliseconds, a typical survey optimized for MSPs
employs several thousand frequency channels
and samples data every 100~$\mu$s or less.
This pushes MSP searches to very high data rates and well
into the realm of the National Science Foundation's new buzzword:
``big data''.  Furthermore, about three quarters of all MSPs are found
in binary systems, whereas most long-period pulsars are isolated (see Fig.~\ref{fig:evolution}a).
Acceleration in a binary system induces a Doppler shift in the
observed pulse period that would render many systems undetectable
without specialized (and computationally intensive) search techniques.
Systems in which the companion eclipses the pulsar for large fractions
of time may require multiple observations to detect.  All of these
factors make high-performance computing a must for modern pulsar
surveys.  The scientific payoff is well worth the effort, however, and
indeed is absolutely necessary in the era of pulsar timing arrays.  A
successful detection of gravitational waves will require many
ultra-high precision MSPs, ideally distributed isotropically across
the sky.  The optimal MSP for a PTA will be bright and have narrow
pulse profile features, allowing for precise pulse time of arrival
(TOA) measurements.  It will furthermore be free from difficult to
model binary effects (such as eclipses and interactions with
intra-binary gas) and will suffer from a minimum of ISM effects (see the
article in this focus issue by D. Stinebring for an in-depth discussion
of ISM effects), both of which can lead to large residuals in pulsar
timing models.  Additionally, the optimal MSP will have minimum noise
that is intrinsic to the pulsar, such as variability in emission
from its magnetosphere, an effect known as phase jitter~\citep{cs10,ovh+11} For
an overview of the noise processes intrinsic to pulsars, see the article
in this focus issue by J. M. Cordes. Finding such pulsars is now a primary
motivation for large-area and/or targeted pulsar surveys at nearly all major
radio observatories.

The evolutionary pathways that lead to the variety of observed MSPs
are not fully understood.  An excellent starting point is the
cartoon showing the formation of the various systems
 in Fig.~\ref{fig:evolution}b.  Starting with a
binary system, a neutron star is formed during the supernova explosion
of the usually initially more massive star. Most binaries do not
survive this event, and over 90\% \citep{py98}
are disrupted due to either catastrophic mass
loss~\citep{hil83} and/or natal kicks imparted to the neutron
star~\citep{bai89}.  For binaries which survive, and where the
companion evolves into a red giant, the old spun-down neutron star can
be revived as a pulsar by accreting matter from its companion,
spinning it up to shorter periods~\citep{acrs82}. The term ``recycled
pulsar'' is used to describe such objects. During accretion, X-rays
produced by the frictional heating of in-falling matter onto the
neutron star make such systems visible as X-ray binaries. For an
overview of X-ray binaries, see e.g.~\cite{bv91}.

Two classes of X-ray binaries exist that are relevant to recycled 
pulsars: neutron
stars with high-mass or low-mass companions. In a high-mass X-ray
binary, the companion is massive enough that it evolves on a
$10^{6-7}$~yr timescale before exploding as a supernova, producing a
second neutron star. For binaries which survive
the explosion, the result is a double neutron star binary. At least
nine such systems are currently known~\citep{lor08}.  Most relevant to
the formation of MSPs are low-mass X-ray binary systems (LMXBs) where the
companion evolves and transfers matter onto the neutron star on a much
longer timescale (of order $10^8$~yr or more), spinning it up to
periods as short as a few ms~\citep{acrs82}. Tidal forces during the
accretion process serve to circularize the orbit.  During this
spin-up phase, the secondary sheds its outer layers to become a white
dwarf in a circular orbit around a rapidly spinning MSP.  This theoretical
description connecting LMXBs to the formation of new MSPs has been confirmed
observationally in recent years in the detection of an accretion disk around
PSR J1023+0038~\citep{asr+09} and even more recently the connection of the
LMXB IGR J18245-2452 and PSR J1824-2452~\citep{pfb+13}.

\section{Searching the Sky for New MSPs}\label{sec:searching}

\begin{figure}[h!]
\includegraphics[width=\textwidth]{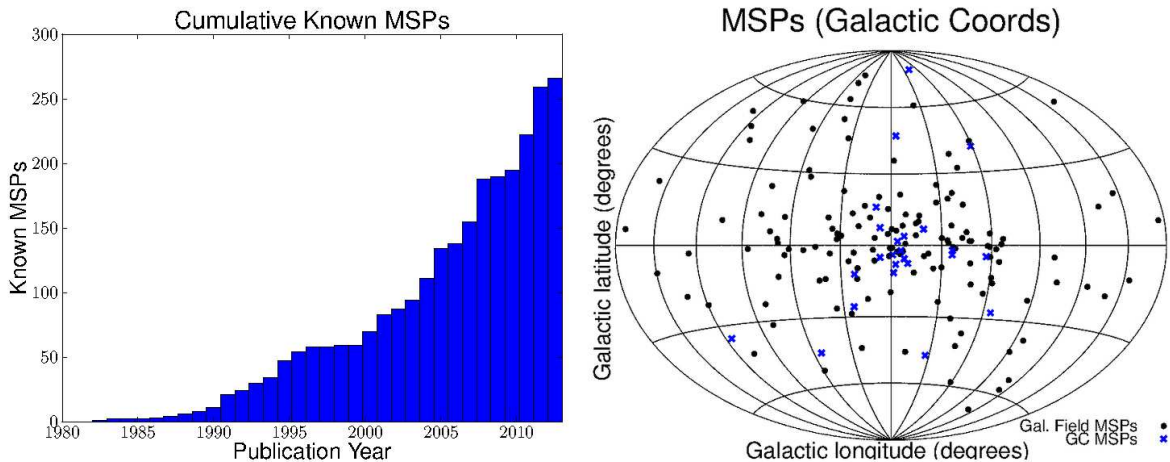}
\begin{center}
\caption[]{Left: The number of MSPs from the ATNF pulsar catalog (see Appendix A) in the Galactic
field and globular clusters as a function of publication date. Right: MSPs in the
Galactic field (black dots) and in globular clusters (blue X's) shown in Galactic coordinates. The
center of the image corresponds to the location of the Galactic center and the x-axis represents
the Galactic plane.}
\label{fig:MSPsvdate}
\end{center}
\end{figure}

Fig.~\ref{fig:MSPsvdate} summarizes the progress in searching for MSPs
since their discovery in 1982~\citep{bkh+82}. As can be seen, thanks to a large number of
different surveys being carried out with different telescopes (see Subsections ~\ref{subsec:target}
and ~\ref{subsec:surveys} below), we are currently enjoying a burst in the number of systems being found. 

There are two general methods for searching for new pulsars. One
is to do targeted searches of known objects such as globular clusters.
In recent years, searches of sources identified,
using the Fermi Gamma-Ray Space Telescope, as point sources with pulsar
characteristics have lead to the discovery of many new MSPs. The second
method for finding new pulsars is to systematically search
large regions of the sky. We will describe each of these approaches
in more detail below.

\subsection{Targeted Searches}\label{subsec:target}

Globular clusters are excellent targets for finding new MSPs. Due to their
increased stellar density in comparison to the Galactic field, globular
clusters are environments in which interactions between stars, such as those
required to produce MSPs, are more likely to occur. In addition, the
high probability of exchange interactions \citep{sp90,ps91}
means that the observed population of MSPs in clusters has a 
significant fraction of ``exotic systems'' with high orbital eccentricities
\citep{dbl+93},
short orbital periods \citep{clf+00} and large inferred masses \cite{fwvh08}.
To date, a total of 
144 pulsars have been found in 28 cluster
\footnote{See www.naic.edu/$\sim$pfreire/GCpsr.html for more information.}. All but four of these pulsars
show characteristics akin to recycled pulsars and 129 of the currently
known pulsars have $P<20$~ms. Searches of globular clusters
have long been fruitful for finding new MSPs
\citep{mlr+91,bbl+94,dlm+01,rhs+05},
but from a PTA standpoint
they are of limited use.  Acceleration in the globular cluster potential,
as well as acceleration and jerk from nearby stars, leads to large timing
residuals on long timescales.  Globular clusters are also typically a factor
of a few more distant than typical PTA MSPs, and hence typically an order
of magnitude fainter. 

Targeted searches of bright \emph{Fermi} point sources have been
amazingly successful at identifying new MSPs. To date, more than
50 MSPs have been discovered that are coincident with \emph{Fermi} point
sources. Many of these are ``black widow'' or ``redback'' systems
that often eclipse due to excessive material surrounding the companion and
show significant and, from the point of view of PTAs, undesirable intra-binary
effects.  For further information on such systems, see \cite{fre04}, \cite{rob11},
and references therein. Despite the number of ``black widow'' and ``redback''
systems in the \emph{Fermi}-discovered set of MSPs, at least 10 are now
being timed regularly by the various PTA projects. A recent review of the
\emph{Fermi} searches can be found in Ray et al.~(2012). \nocite{rap+12}

\subsection{Large Area Surveys}\label{subsec:surveys}

Although targeted searches have been successful in finding new MSPs,
large area surveys are required in order to find the MSPs in
the Galactic field which are not necessarily strong gamma-ray emitters. 
Following initial searches in the 1980s with relatively poor sensitivity,
large-area surveys have undergone two major renaissance periods.
The first of these took place during the 1990s following Wolszczan's
discovery of two recycled pulsars at high Galactic latitudes in an
Arecibo drift-scan survey \citep{wol91a}. Subsequently,
a prescient paper by Johnston \& Bailes (1991) 
demonstrated that the local population of MSPs
revealed by all-sky surveys at $\sim 0.4$~GHz
should be largely isotropic. 
This work
inspired a number of 400~MHz pulsar surveys during the 1990s which led
to a sample of about 30 MSPs by the end of the decade. The main contributions
made were at Parkes where a 436 MHz Survey of the southern sky resulted in the discovery of 17 new MSPs
\citep{mld+96}, and at Arecibo where a number of groups surveyed 
significant portions of the Arecibo visible sky and found a similar number of MSPs.
For an excellent review of these searches, see Camilo (1995, 1999). \nocite{cam95,cam99}

The second renaissance period began during the late 1990s with the
emergence of the most prolific pulsar survey so far, the Parkes
Multibeam Pulsar Survey (PMPS).  The PMPS made use of the 20-cm
(L-band) multibeam system on the Parkes Telescope to survey the
Galactic plane for new pulsars using 13 independent beams at a time
\citep{mlc+01}. So far, well over 1000 pulsars have been discovered by
this system and the original PMPS survey of the Galactic plane
($|b|<5^{\circ}$) carried out in the early 2000s
\citep{mlc+01,mhl+02,kbm+03,hfs+04,fsk+04,lfl+06}
has discovered around 800 new pulsars including 30 new
MSPs. Additional discoveries are still being made by groups
reprocessing the data \citep{eklk13,kek+13,mlb+12}.  Following the
success of the PMPS, surveys extending the surveyed area to
intermediate and high latitudes were also performed using the Parkes
multibeam L-band feed. These surveys are the Swinburne Intermediate
Latitude Survey \citep{ebvb01} and the Swinburne High Latitude Survey
\citep{jbo+09}. These two surveys discovered 8 and 5 new MSPs,
respectively.

Inspired by the success of the PMPS surveys, a 7-beam L-band system
was commissioned at Arecibo in 2004 and has been used for pulsar and
neutral hydrogen surveys ever since. The Pulsar Arecibo L-band Feed
Array (PALFA) survey covers the region of the Galaxy with galactic
latitude less than $\pm5^\circ$ in the galactic longitude ranges
$32^\circ<\ell<77^\circ$ and $168^\circ<\ell<214^\circ$. Initial data
were taken using a data acquisition system with 100 MHz bandwidth
which has subsequently been upgraded to sample the full 322~MHz band
from the receiver. To date, the PALFA survey has discovered 116
pulsars, with 17 of these being new MSPs.

With most of the pulsar searching efforts at Arecibo and Parkes being
devoted to L-band multibeam systems, at Green Bank, an opportunity to
return to drift-scan searching arose when the GBT was closed to allow
repair of its azimuth track during summer 2007.  Drift-scan observations
at 350~MHz carried out during this period covered over 10000~deg$^2$
in the declination ranges $-7.7^\circ \leq \delta \leq 38.4^\circ$ and
$-20.7^\circ \leq \delta \leq 38.4^\circ$. Further details of the
survey coverage, data processing, and sensitivity can be found in
\citet{blr+13} and \citet{lbr+13}.  Data processing for the Drift-scan
survey is now complete, and 35 pulsars have been discovered, including
7 MSPs and recycled pulsars.  Twenty-four pulsars from early data
processing are presented in \citet{blr+13} and
\citet{lbr+13} along with complete timing solutions.  An additional 11
pulsars have been discovered since this first round of detailed
follow-up and are still being studied.

\subsubsection{Current Surveys}\label{subsubsec:cursurveys} 

The PTA experiments and advancements in data recording have driven the
development of many large pulsar surveys which are currently underway.

The Arecibo Observatory 327 MHz Drift Scan (AO327) Survey \citep{dsm+13}
is an ongoing survey using the Arecibo
Observatory, which is intended to search the entire sky visible by the
telescope for new radio pulsars at 327 MHz. In this survey, the
telescope is fixed at a particular azimuthal and zenith position while
the rotation of the Earth moves the sky overhead. Due to the nature of
this survey (it is sometimes performed when telescope pointing is not
functional), it covers right ascension ranges spread throughout the
declination range $-1^\circ<\delta<38^\circ$.  To date, the survey has
discovered 22 new pulsars including 3 MSPs.

The Green Bank North Celestial Cap (GBNCC) survey~(Stovall et al.~2013,
in preparation) is the successor
to the aforementioned GBT 350-MHz drift scan survey and is
also carried out at $350\; $MHz, giving it excellent sensitivity to
nearby, steep spectrum pulsars.  It uses twice the bandwidth of the former
survey, 120-s pointed observations, and the
newer Green Bank Ultimate Pulsar Processor back-end \citep{drd+08}.
The science
goals are the same, but with a particular emphasis on northern
declinations, where there are fewer high-precisions MSPs known,
especially in the first stage of the survey.  This is important for
increasing the number of wide-separation baselines in PTAs and also
probes a region of the Galaxy that has not been studied in as much
detail as the Galactic plane.  Stage I of the survey covered the north
celestial cap ($\delta > 38^\circ$) and data taking was completed in
2011.  The second stage, which covers the remaining GBT visible sky,
is currently underway.
Data processing is being carried out at the Texas Advanced Computing
Center and the Guillimin supercomputer operated by CLUMEQ.
To date, the survey has discovered 62 new pulsars, including 9 new MSPs.
Further analysis of candidates from the GBNCC survey,
as well as ongoing data-taking and processing, will undoubtedly result
in the discovery of many more pulsars.

The High Time Resolution Universe (HTRU) pulsar survey~\citep{kjv+10} 
is currently 
underway at the Parkes Telescope, using the
same multibeam system as was used by the PMPS. Data acquisition is being 
carried out using updated spectrometers which provide order-of-magnitude
increases in time and frequency resolution over the previous generation
of Parkes multibeam surveys. The HTRU surveys are divided 
into three sky areas; the low, intermediate, and high Galactic latitude regions.
These three regions have integration times of 4300 s, 540 s, and 270 s, respectively. 
To date, the HTRU survey has resulted in the discovery of close to 150 pulsars, including
about 30 new MSPs. 
A similar survey (HTRU-N) is also being conducted in the northern sky using a new seven beam system at the
Effelsberg radio telescope~\citep{bck+13}.

In recent years, low frequency observatories such as the Low Frequency Array~\citep[LOFAR,][]{sha+11},
the Long Wavelength Array~\citep[LWA,][]{tek+12}, and the Murchison Widefield Array~\citep[MWA,][]{bck+13b}
have begun to become operational. These instruments are beginning to
be used for large area pulsar surveys, which may lead to the detection of new MSPs. However, searches conducted
at these low frequencies ($\sim$100 MHz) are more challenging due to the effects of the interstellar medium. The effects
of dispersion are so strong below 100 MHz, that searches for MSPs at even moderate DMs require high frequency resolution to adequately
account for dispersion broadening effects.
Additionally, the
detectable MSP population is reduced at these frequencies due to scattering by the ISM.

In summary, the various technical developments over the past decade have
led to the discoveries of over 150 MSPs in the Galactic disk. In addition
to finding MSPs in large numbers, and therefore increasing the chances
of uncovering one that times well, the same developments have led to a 
dramatic improvement in time and frequency resolution available for
follow-up timing experiments. MSPs can now be routinely timed to $\mu$s
precision or better with 100-m class telescopes. A decade ago, such a 
statement could only be made for pulsars timed with the large collecting
area (and, hence, excellent signal-to-noise ratio) of the 305-m Arecibo
telescope \citep{cam99}.

\section{Search Method}\label{sec:searchmethod}

The pulsar surveys and targeted searches use slightly varying techniques for discovering
new pulsars. Here we discuss the basic methods performed. Information on other search methods
as well as a more in-depth discussion of the methods described below can be found in \cite{lk05}.
Standard software packages capable of performing many of the following tasks are available
online. These packages include {\tt PRESTO}, {\tt SIGPROC}, and {\tt PSRCHIVE}. Links to these
packages are included in Appendix A.

\subsection{RFI Excision}\label{subsec:rfi}

One of the challenges faced by pulsar searches is the increasing levels of radio frequency
interference (RFI) created by a technologically growing world. Pulsar searchers use several
techniques to mitigate the effects of RFI. There are two main types of RFI, strong bursts and
low-level, continuous signals. In order to remove bursts of RFI, the data are generally analyzed in
sections to identify periods of time with increased power, or specific frequency channels which contain
significantly more power than the others for the same observation. The time periods and frequency channels
with too much power are then identified and either not used in analysis or the data for that time period/frequency
channel are replaced with the mean of the observation. The removal of low-level, continuous signals is
often done by removing power from specific Fourier bins, which are known to generally have increased
power due to RFI.

\subsection{De-dispersion}\label{subsec:dedisp}

Prior to searching data for pulsars, we must remove the effects of dispersion, which will smear the signal
out as a function of frequency. In order to remove this effect, the typical method for pulsar searching
is to obtain data which is divided into a number of frequency channels. The number of frequency channels
depends on observing frequency and is chosen such that the dispersive smearing time over individual channels
is much less than the period of the pulsars you wish to find out to some DM. When performing the searches, we
do not know the DM of our pulsars, so we must search over a wide range of trial DMs. The range and step sizes
used are dependent on the observing frequency, bandwidth, number of frequency channels, 
direction of search, and
available computational resources. A typical method is to create a set of DMs such that smearing due
to DM error is less than the smearing within the frequency channels. Also, when the
smearing time within the frequency channels is $2^N$ times the sample time, the time series
is down-sampled by a factor of $2^N$.

\subsection{Search Algorithms}\label{subsec:search}

There are various search methods for finding new MSPs. The most common is to perform an FFT and
then incoherently sum harmonics~\citep{th69}. Binary pulsars' periods will change due to the
Dopplar shift from orbital motion. If the period change is significant over the length of an
observation, then the pulsars' signal will be smeared across multiple Fourier bins. In order
to mitigate this smearing, techniques assuming a constant acceleration such as the correlation
method~\citep{rgh+01} or the 
stack/slide technique~\citep{fsk+04} are often used. These techniques
are effective for orbital periods much larger than the observation time. In the case of
the current large sky surveys, which have observing times of a few minutes, these linear
acceleration techniques are adequate.

\subsection{Sifting and Folding}\label{subsec:fold}

\begin{figure}
  \includegraphics[width=0.9\textwidth]{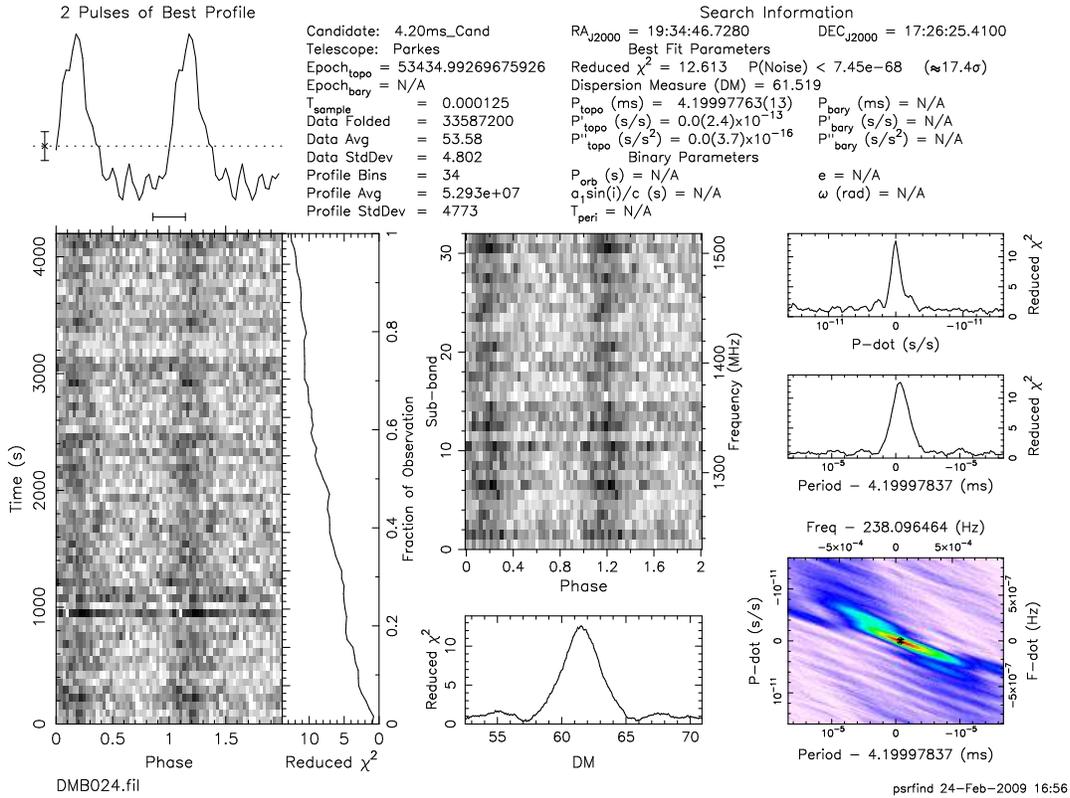}
  \caption{\label{fig:prepfold}
Example periodicity search output plot
showing the discovery observation of the 4.2-ms pulsar J1935+1726
\citep{lcm13}
 folded in time
(lower left) and radio frequency (upper center) as well as the integrated
pulse profile (upper left) and optimal DM search (lower center). 
The statistical significance of the signal in each of these diagrams
is measured in terms of the
reduced $\chi^2$ square value computed from the integrated pulse profiles.
A $\chi^2$ value close to unity would be found for a profile that is consistent
with Gaussian random noise. 
}
\end{figure}

After each of the de-dispersed timeseries have been searched, the
resulting candidate signals must be sifted through in order to
determine the likely pulsar candidates and remove ones which are
unlikely to be real. Typical tactics used are to remove candidates
which have periods of known RFI, do not appear at contiguous trial
DMs, or are detected at only a single DM value.  The candidates which
make it through the sifting process are then made into diagnostic
plots of the form shown in Fig.~\ref{fig:prepfold}.  In some cases,
the resulting number of candidates can be too large for all candidates
to be plotted, in which case only the top candidates are plotted.

\subsection{Candidate Analysis}\label{subsec:candidates}

Over the years, the instruments used to perform pulsar searches have improved drastically
in frequency coverage, frequency resolution, and time resolution. One of the effects of
these improvements is a drastic increase in the number of candidates generated by a search.
This increase in candidates generated by pulsar surveys must be handled in some way by each
of the survey groups. In recent years, there have been two major approaches to dealing with
the large number of pulsar candidates. One is to develop automatic algorithms which are used
to classify the candidates in such a way that real pulsars are distingushed from the RFI
and noise. The other is to increase the number of people analyzing the data by training
high school and undergraduate students to recognize pulsar signals. These two
methods are described in detail below.

\subsection{Automatic Search Algorithms}\label{subsubsec:auto} 

A majority of the search algorithms currently in use and being
developed for distinguishing real pulsars from RFI and noise are based
on what humans typically look for in the diagnostic plots created for
each candidate. Some basic heuristics which are looked for in a
candidate are: whether the signal is broadband, is detected throughout
the observation, has a similar pulse profile to most known pulsars,
has a DM curve shaped as expected, etc. Two tools which calculate
these heuristics and provide methods for initial sorting are the
JReaper tool~\citep{kel+09} and the PEACE algorithm~\citep{lsj+13}.
Other techniques that take the heuristics calculated by tools like
JReaper and PEACE and pass them to machine learning algorithms have
also been developed or are currently being developed.  The HTRU survey
applied an artificial neural network which recovered 92\% of known
pulsars from $\sim$2.5 million candidates~\citep{emk+10}.

\subsection{Who searches for pulsars?}\label{subsubsec:students} 

Pulsar searching has traditionally been the purview of graduate
students as part of their thesis work\footnote{All three of the
  authors of the current paper spent a substantial fraction of their
  time sifting through pulsar search output!}. In recent years,
however, a number of groups have involved other groups through
innovative outreach projects.  In Australia the Pulse@Parkes project
\citep{hlc+09} gives high-school students the opportunity to
carry out regular timing observations of pulsars using the Parkes
radio telescope and learn about their properties. In Germany, the
Einstein@Home project \citep{akc+13} allows citizen scientists across the
world the opportunity to participate in the discovery process by
donating idle cycles on their home computers. Based on the Seti@Home
infrastructure, Einstein@Home volunteers have so far discovered over
20 pulsars in searches of Arecibo and Parkes survey 
data~\citep{kla+11,kek+13}.

In the US, two main pulsar searching outreach projects exist.  The
Arecibo Remote Command Center (ARCC) is a group of high school,
undergraduate, and graduate students who work with university
professors and local high school teachers in a program to detect new
radio pulsars. The group was formed to get undergraduate students
involved in research early in their careers, assist in controlling
radio telescopes during pulsar survey observations, and to create a
large group of people to inspect the candidates from the pulsar
surveys described in Sec.~\ref{subsec:surveys}.  Students in this
program have found a total of 46 pulsars over the past 3 years.
  There are currently
ARCC centers at the University of Texas at Brownsville (UTB) and the
University of Wisconsin-Milwaukee (UWM).  Another outreach effort
which focuses on high-school and middle-school students is the Pulsar
Search Collaboratory (PSC), an NSF-funded project involving students
and teachers analysing 2800~deg$^2$ of the GBT drift-scan survey in
partnership with the National Radio Astronomy Observatory and West
Virginia University.  Since the PSC began in 2008, nearly 800 students
and 100 teachers have been involved in over 90 schools spread across 18
states in America.  To date, students have inspected over 1.5 million
search diagnostic plots and found a total of 6 new pulsars and
identified previously known pulsars at a rate that is consistent with
professional astronomers \citep{rsm+13}.  Evaluation studies
indicate that the PSC significantly increases student interest in
science, engineering and computer science careers.

\section{The MSP population}

Understanding the origin and evolution of MSPs has provided 
a wealth of information and interesting puzzles over the years.
One of the first efforts to quantify the MSP population
was the work of \citet{kn88}. With a sample of only three
MSPs, their study was subject to large uncertainties,
but it began a significant discussion on the so-called ``birthrate
problem'' for MSPs. Based on their results Kulkarni \&
Narayan  claimed that the birthrate of MSPs was
substantially greater than that of their proposed progenitors, the
low-mass X-ray binaries. Although this problem was alleviated as
better constraints became available from larger 
samples \citep{lor08}, very recently MSP population study
\citep{lbb+13} suggested that the birthrate problem persists.

Following the 400~MHz MSPs surveys that took place during the 1990s,
studies of the scale height, velocity distribution and
luminosity function were performed \citep{lor95,cc97,lml+98}
and it was found that the local (within a few kpc)
MSP population potentially observable was comparable in
size to the equivalent population of normal pulsars. One conclusion
from these studies is that the populations of millisecond and normal
pulsars are consistent with a single velocity distribution applied to
all neutron stars at birth \citep{tb96}.

We are now in an era where the sample
of MSPs is numerous enough to gain further insights
into the population. As recently shown \citep{lor13}, the MSP
population can be described by a model in which the population
of 30,000 potential observable MSPs
has a luminosity function that is log-normal \citep[consistent with the
normal pulsars and recycled pulsars in globular clusters]{fk06,blc11},
an exponential scale height
of 500~pc and a Gaussian radial distribution with a standard deviation of 7.5~kpc.
Further work in this area is needed to quantify more subtle effects.
 Of particular interest are studies of the
motion of MSPs in the $P-\dot{P}$ diagram and the 
relationship to the low-mass X-ray binary population. Additional
work, along the lines of the population syntheses carried
out by \citet{sgh07}, seems to be the next logical step. Significant
progress is now being made in modeling the binary evolutionary steps and
predicting distributions for orbital parameters for the binary population
\citep[see, for example,][]{bkr+08}. Combining all these
elements into an all-encompassing synthesis of the MSP
population which accounts (as far as possible) for the observational
selection effects is now a major goal of future studies. As part of this effort,
a significant problem is to incorporate the gamma-ray/radio-selected sample of MSPs
revealed by {\it Fermi} \citep{rap+12}. Are these pulsars more energetic
than other MSPs? Are the spin periods of these MSPs shorter than the rest of the
population? The answers are currently not clear and a careful study of the
selection effects impacting this sample should now be undertaken in
order to fully understand the impact of these discoveries on our
knowledge of the MSP population.

\section{Future surveys and prospects}\label{sec:conc}

The currently ongoing pulsar surveys are close to the most sensitive
surveys possible with the current set of telescopes. So, in order to
make significant leaps in sensitivity, instruments with larger
collecting areas must be built.  The next generation of telescopes are
currently being planned and in some cases constructed. In the near
future, MeerKAT and FAST will begin to come online. MeerKAT, an array
of sixty-four 13.5 m dishes, will have about the equivalent
sensitivity of a 100-m telescope. This will be an increase of about
2.5 over the existing telescopes in the southern hemisphere. The Five
hundred meter Aperture Spherical Telescope (FAST) telescope is an
Arecibo-like telescope which will have a diameter of 500 m. It will
provide an increase in sensitivity of about a factor of 2 in regions
of the sky visible by Arecibo and an increase by a factor of ten in
other areas of the sky.

In the long term, the initial phase
of the Square Kilometer Array (SKA) is expected to be
built and will provide an increased sensitivity of about a factor of
1.3 in regions overlapping with FAST and a factor of more than 10 times
in other regions of the sky. The field-of-view of the SKA will be significantly
larger than that of the FAST telescope and therefore will have a much
faster survey speed.

Other than FAST, the current trend in radio telescope development is
to build large numbers of smaller antennas operating together as an
array. These arrays alter the way that pulsar surveys are done~\citep{sha+11,coe13}. The
time required to survey the entire sky is significantly reduced, since
the beams of individual telescopes are quite large and multiple sky
beams can be taken at once. However, the computational requirements
and amount of data will also be significantly larger.

\section*{Acknowledgements}

DRL was supported by Oxford Astrophysics while on sabbatical leave.

\appendix

\section{Useful resources}

\begin{table}[h!]
\begin{small}
\begin{tabular}{ll}
URL & Description \\
\hline
www.atnf.csiro.au/people/pulsar/psrcat & ATNF pulsar catalog \\
astro.phys.wvu.edu/GalacticMSPs & Galactic MSPs \\
www.naic.edu/$\sim$pfreire/GCpsr.html & Pulsars in globular clusters \\
www.naic.edu/$\sim$palfa/newpulsars & PALFA survey discoveries \\
astro.phys.wvu.edu/pmps & PMPS reprocessing at WVU \\
albert.phys.uwm.edu/radiopulsar/html/PMPS\_discoveries & PMPS (Einstein@Home) \\
astro.phys.wvu.edu/GBTdrift350 & GBT 350~MHz driftscan\\
arcc.phys.utb.edu/gbncc & GBNCC\\
www.naic.edu/$\sim$deneva/drift-search & Arecibo 327~MHz driftscan\\
www.astron.nl/pulsars/lofar/surveys/lotas & LOFAR Pilot Pulsar Survey \\
www.pulsarsearchcollaboratory.com& Pulsar Search Collaboratory\\
arcc.phys.utb.edu & Arecibo Remote Command Center\\
outreach.atnf.csiro.au/education/pulseatparkes & Pulse@Parkes\\
www.pulsarastronomy.net/wiki/Software/PulsarHunter & JREAPER\\
psrchive.sourceforge.net & PSRCHIVE\\
http://www.cv.nrao.edu/$\sim$sransom/presto & PRESTO\\
sigproc.sourceforge.net & SIGPROC\\
\hline
\end{tabular}
\end{small}
\end{table}


\begin{thebibliography}{81}
\expandafter\ifx\csname natexlab\endcsname\relax\def\natexlab#1{#1}\fi

\bibitem[{{Abdo} {et~al}\mbox{.}(2009){Abdo}, {Ackermann}, {Ajello}, {Atwood},
  {Axelsson}, {Baldini}, {Ballet}, {Barbiellini}, {Baring}, {Bastieri},
  {Baughman}, {Bechtol}, {Bellazzini}, {Berenji}, {Bignami}, {Blandford},
  {Bloom}, {Bonamente}, {Borgland}, {Bregeon}, {Brez}, {Brigida}, {Bruel},
  {Burnett}, {Caliandro}, {Cameron}, {Camilo}, {Caraveo}, {Carlson},
  {Casandjian}, {Cecchi}, {{\c C}elik}, {Charles}, {Chekhtman}, {Cheung},
  {Chiang}, {Ciprini}, {Claus}, {Cognard}, {Cohen-Tanugi}, {Cominsky},
  {Conrad}, {Corbet}, {Cutini}, {Dermer}, {Desvignes}, {de Angelis}, {de Luca},
  {de Palma}, {Digel}, {Dormody}, {do Couto e Silva}, {Drell}, {Dubois},
  {Dumora}, {Edmonds}, {Farnier}, {Favuzzi}, {Fegan}, {Focke}, {Frailis},
  {Freire}, {Fukazawa}, {Funk}, {Fusco}, {Gargano}, {Gasparrini}, {Gehrels},
  {Germani}, {Giebels}, {Giglietto}, {Giordano}, {Glanzman}, {Godfrey},
  {Grenier}, {Grondin}, {Grove}, {Guillemot}, {Guiriec}, {Hanabata}, {Harding},
  {Hayashida}, {Hays}, {Hobbs}, {Hughes}, {J{\'o}hannesson}, {Johnson},
  {Johnson}, {Johnson}, {Johnson}, {Johnston}, {Kamae}, {Katagiri}, {Kataoka},
  {Kawai}, {Kerr}, {Kn{\"o}dlseder}, {Kocian}, {Kramer}, {Kuss}, {Lande},
  {Latronico}, {Lemoine-Goumard}, {Longo}, {Loparco}, {Lott}, {Lovellette},
  {Lubrano}, {Madejski}, {Makeev}, {Manchester}, {Marelli}, {Mazziotta},
  {McConville}, {McEnery}, {McLaughlin}, {Meurer}, {Michelson}, {Mitthumsiri},
  {Mizuno}, {Moiseev}, {Monte}, {Monzani}, {Morselli}, {Moskalenko}, {Murgia},
  {Nolan}, {Norris}, {Nuss}, {Ohsugi}, {Omodei}, {Orlando}, {Ormes}, {Paneque},
  {Panetta}, {Parent}, {Pelassa}, {Pepe}, {Pesce-Rollins}, {Piron}, {Porter},
  {Rain{\`o}}, {Rando}, {Ransom}, {Ray}, {Razzano}, {Rea}, {Reimer}, {Reimer},
  {Reposeur}, {Ritz}, {Rochester}, {Rodriguez}, {Romani}, {Roth}, {Ryde},
  {Sadrozinski}, {Sanchez}, {Sander}, {Saz Parkinson}, {Scargle}, {Schalk},
  {Sgr{\`o}}, {Siskind}, {Smith}, {Smith}, {Spandre}, {Spinelli}, {Stappers},
  {Starck}, {Striani}, {Strickman}, {Suson}, {Tajima}, {Takahashi}, {Tanaka},
  {Thayer}, {Thayer}, {Theureau}, {Thompson}, {Thorsett}, {Tibaldo}, {Torres},
  {Tosti}, {Tramacere}, {Uchiyama}, {Usher}, {Van Etten}, {Vasileiou},
  {Venter}, {Vilchez}, {Vitale}, {Waite}, {Wallace}, {Wang}, {Watters}, {Webb},
  {Weltevrede}, {Winer}, {Wood}, {Ylinen}, \& {Ziegler}}]{aaa+09}
{Abdo} A.~A. {et~al.}, 2009, Science, 325, 848

\bibitem[{{Allen} {et~al}\mbox{.}(2013){Allen}, {Knispel}, {Cordes}, {Deneva},
  {Hessels}, {Anderson}, {Aulbert}, {Bock}, {Brazier}, {Chatterjee},
  {Demorest}, {Eggenstein}, {Fehrmann}, {Gotthelf}, {Hammer}, {Kaspi},
  {Kramer}, {Lyne}, {Machenschalk}, {McLaughlin}, {Messenger}, {Pletsch},
  {Ransom}, {Stairs}, {Stappers}, {Bhat}, {Bogdanov}, {Camilo}, {Champion},
  {Crawford}, {Desvignes}, {Freire}, {Heald}, {Jenet}, {Lazarus}, {Lee}, {van
  Leeuwen}, {Lynch}, {Papa}, {Prix}, {Rosen}, {Scholz}, {Siemens}, {Stovall},
  {Venkataraman}, \& {Zhu}}]{akc+13}
{Allen} B. {et~al.}, 2013, ArXiv e-prints

\bibitem[{Alpar {et~al}\mbox{.}(1982)Alpar, Cheng, Ruderman, \&
  Shaham}]{acrs82}
Alpar M.~A., Cheng A.~F., Ruderman M.~A., Shaham J., 1982, Nature, 300, 728

\bibitem[{{Antoniadis} {et~al}\mbox{.}(2013){Antoniadis}, {Freire}, {Wex},
  {Tauris}, {Lynch}, {van Kerkwijk}, {Kramer}, {Bassa}, {Dhillon}, {Driebe},
  {Hessels}, {Kaspi}, {Kondratiev}, {Langer}, {Marsh}, {McLaughlin},
  {Pennucci}, {Ransom}, {Stairs}, {van Leeuwen}, {Verbiest}, \&
  {Whelan}}]{afw+13}
{Antoniadis} J. {et~al.}, 2013, Science, 340, 448

\bibitem[{{Archibald} {et~al}\mbox{.}(2009){Archibald}, {Stairs}, {Ransom},
  {Kaspi}, {Kondratiev}, {Lorimer}, {McLaughlin}, {Boyles}, {Hessels}, {Lynch},
  {van Leeuwen}, {Roberts}, {Jenet}, {Champion}, {Rosen}, {Barlow}, {Dunlap},
  \& {Remillard}}]{asr+09}
{Archibald} A.~M. {et~al.}, 2009, Science, 324, 1411

\bibitem[{Backer {et~al}\mbox{.}(1982)Backer, Kulkarni, Heiles, Davis, \&
  Goss}]{bkh+82}
Backer D.~C., Kulkarni S.~R., Heiles C., Davis M.~M., Goss W.~M., 1982, Nature,
  300, 615

\bibitem[{{Bagchi}, {Lorimer} \& {Chennamangalam}(2011){Bagchi}, {Lorimer}, \&
  {Chennamangalam}}]{blc11}
{Bagchi} M., {Lorimer} D.~R., {Chennamangalam} J., 2011, MNRAS, 418, 477

\bibitem[{Bailes(1989)}]{bai89}
Bailes M., 1989, ApJ, 342, 917

\bibitem[{Barr {et~al}\mbox{.}(2013)Barr, Champion, Kramer, Eatough, Freire,
  Karuppusamy, Lee, Verbiest, Bassa, Lyne, Stappers, Lorimer, \&
  Klein}]{bck+13}
Barr E.~D. {et~al.}, 2013, MNRAS, in press

\bibitem[{{Belczynski} {et~al}\mbox{.}(2008){Belczynski}, {Kalogera}, {Rasio},
  {Taam}, {Zezas}, {Bulik}, {Maccarone}, \& {Ivanova}}]{bkr+08}
{Belczynski} K., {Kalogera} V., {Rasio} F.~A., {Taam} R.~E., {Zezas} A.,
  {Bulik} T., {Maccarone} T.~J., {Ivanova} N., 2008, ApJS, 174, 223

\bibitem[{{Bhat}, {Bailes} \& {Verbiest}(2008){Bhat}, {Bailes}, \&
  {Verbiest}}]{bbv08}
{Bhat} N.~D.~R., {Bailes} M., {Verbiest} J.~P.~W., 2008, Phys. Rev. D, 77,
  124017

\bibitem[{Bhattacharya \& {van den Heuvel}(1991)}]{bv91}
Bhattacharya D., {van den Heuvel} E. P.~J., 1991, Phys. Rep., 203, 1

\bibitem[{Biggs {et~al}\mbox{.}(1994)Biggs, Bailes, Lyne, Goss, \&
  Fruchter}]{bbl+94}
Biggs J.~D., Bailes M., Lyne A.~G., Goss W.~M., Fruchter A.~S., 1994, MNRAS,
  267, 125

\bibitem[{{Bowman} {et~al}\mbox{.}(2013){Bowman}, {Cairns}, {Kaplan}, {Murphy},
  {Oberoi}, {Staveley-Smith}, {Arcus}, {Barnes}, {Bernardi}, {Briggs}, {Brown},
  {Bunton}, {Burgasser}, {Cappallo}, {Chatterjee}, {Corey}, {Coster},
  {Deshpande}, {deSouza}, {Emrich}, {Erickson}, {Goeke}, {Gaensler},
  {Greenhill}, {Harvey-Smith}, {Hazelton}, {Herne}, {Hewitt},
  {Johnston-Hollitt}, {Kasper}, {Kincaid}, {Koenig}, {Kratzenberg}, {Lonsdale},
  {Lynch}, {Matthews}, {McWhirter}, {Mitchell}, {Morales}, {Morgan}, {Ord},
  {Pathikulangara}, {Prabu}, {Remillard}, {Robishaw}, {Rogers}, {Roshi},
  {Salah}, {Sault}, {Shankar}, {Srivani}, {Stevens}, {Subrahmanyan}, {Tingay},
  {Wayth}, {Waterson}, {Webster}, {Whitney}, {Williams}, {Williams}, \&
  {Wyithe}}]{bck+13b}
{Bowman} J.~D. {et~al.}, 2013, pasa, 30, 31

\bibitem[{{Boyles} {et~al}\mbox{.}(2013){Boyles}, {Lynch}, {Ransom}, {Stairs},
  {Lorimer}, {McLaughlin}, {Hessels}, {Kaspi}, {Kondratiev}, {Archibald},
  {Berndsen}, {Cardoso}, {Cherry}, {Epstein}, {Karako-Argaman}, {McPhee},
  {Pennucci}, {Roberts}, {Stovall}, \& {van Leeuwen}}]{blr+13}
{Boyles} J. {et~al.}, 2013, ApJ, 763, 80

\bibitem[{Camilo(1995)}]{cam95}
Camilo F., 1995, in The Lives of the Neutron Stars ({NATO ASI Series}), Alpar
  A., Kizilo\u{g}lu {\"U}., {van Paradis} J., eds., Kluwer, Dordrecht, pp.
  243--257

\bibitem[{{Camilo}(1999)}]{cam99}
{Camilo} F., 1999, in Pulsar Timing, General Relativity, and the Internal
  Structure of Neutron Stars, Arzoumanian Z., {van der Hooft} F., {van den
  Heuvel} E.~P.~J., eds., North Holland, Amsterdam, pp. 115--124

\bibitem[{{Camilo} {et~al}\mbox{.}(2000){Camilo}, {Lorimer}, {Freire}, {Lyne},
  \& {Manchester}}]{clf+00}
{Camilo} F., {Lorimer} D.~R., {Freire} P., {Lyne} A.~G., {Manchester} R.~N.,
  2000, ApJ, 535, 975

\bibitem[{{Coenen}(2013)}]{coe13}
{Coenen} T., 2013, in IAU Symposium, Vol. 291, IAU Symposium, pp. 229--232

\bibitem[{Cordes \& Chernoff(1997)}]{cc97}
Cordes J.~M., Chernoff D.~F., 1997, ApJ, 482, 971

\bibitem[{{Cordes} \& {Shannon}(2010)}]{cs10}
{Cordes} J.~M., {Shannon} R.~M., 2010, ArXiv e-prints

\bibitem[{D'Amico {et~al}\mbox{.}(1993)D'Amico, Bailes, Lyne, Manchester,
  Johnston, Fruchter, \& Goss}]{dbl+93}
D'Amico N., Bailes M., Lyne A.~G., Manchester R.~N., Johnston S., Fruchter
  A.~S., Goss W.~M., 1993, MNRAS, 260, L7

\bibitem[{D'Amico {et~al}\mbox{.}(2001)D'Amico, Lyne, Manchester, Possenti, \&
  Camilo}]{dlm+01}
D'Amico N., Lyne A.~G., Manchester R.~N., Possenti A., Camilo F., 2001, ApJ,
  548, L171

\bibitem[{{Demorest} {et~al}\mbox{.}(2010){Demorest}, {Pennucci}, {Ransom},
  {Roberts}, \& {Hessels}}]{dpr+10}
{Demorest} P.~B., {Pennucci} T., {Ransom} S.~M., {Roberts} M.~S.~E., {Hessels}
  J.~W.~T., 2010, Nature, 467, 1081

\bibitem[{Deneva {et~al}\mbox{.}(2013)Deneva, Stovall, McLaughlin, Bates,
  Freire, Jenet, \& Bagchi}]{dsm+13}
Deneva J., Stovall K., McLaughlin M., Bates S., Freire P., Jenet F., Bagchi M.,
  2013, ApJ, submitted

\bibitem[{{DuPlain} {et~al}\mbox{.}(2008){DuPlain}, {Ransom}, {Demorest},
  {Brandt}, {Ford}, \& {Shelton}}]{drd+08}
{DuPlain} R., {Ransom} S., {Demorest} P., {Brandt} P., {Ford} J., {Shelton}
  A.~L., 2008, in Society of Photo-Optical Instrumentation Engineers (SPIE)
  Conference Series, Vol. 7019, Society of Photo-Optical Instrumentation
  Engineers (SPIE) Conference Series

\bibitem[{{Eatough} {et~al}\mbox{.}(2013){Eatough}, {Kramer}, {Lyne}, \&
  {Keith}}]{eklk13}
{Eatough} R.~P., {Kramer} M., {Lyne} A.~G., {Keith} M.~J., 2013, ArXiv e-prints

\bibitem[{{Eatough} {et~al}\mbox{.}(2010){Eatough}, {Molkenthin}, {Kramer},
  {Noutsos}, {Keith}, {Stappers}, \& {Lyne}}]{emk+10}
{Eatough} R.~P., {Molkenthin} N., {Kramer} M., {Noutsos} A., {Keith} M.~J.,
  {Stappers} B.~W., {Lyne} A.~G., 2010, MNRAS, 407, 2443

\bibitem[{{Edwards} {et~al}\mbox{.}(2001){Edwards}, {Bailes}, {van Straten}, \&
  {Britton}}]{ebvb01}
{Edwards} R.~T., {Bailes} M., {van Straten} W., {Britton} M.~C., 2001, MNRAS,
  326, 358

\bibitem[{{Faucher-Gigu{\`e}re} \& {Kaspi}(2006)}]{fk06}
{Faucher-Gigu{\`e}re} C.-A., {Kaspi} V.~M., 2006, ApJ, 643, 332

\bibitem[{{Faulkner} {et~al}\mbox{.}(2004){Faulkner}, {Stairs}, {Kramer},
  {Lyne}, {Hobbs}, {Possenti}, {Lorimer}, {Manchester}, {McLaughlin},
  {D'Amico}, {Camilo}, \& {Burgay}}]{fsk+04}
{Faulkner} A.~J. {et~al.}, 2004, MNRAS, 355, 147

\bibitem[{Freire {et~al}\mbox{.}(2008)Freire, Wolszczan, van~den Berg, \&
  Hessels}]{fwvh08}
Freire P., Wolszczan A., van~den Berg M., Hessels J., 2008, ApJ, 679, 1433

\bibitem[{Freire(2005)}]{fre04}
Freire P.~C., 2005, in {Binary Radio Pulsars}, Rasio F., Stairs I.~H., eds.,
  Astronomical Society of the Pacific, San Francisco, pp. 405--417

\bibitem[{{Freire} {et~al}\mbox{.}(2012){Freire}, {Wex}, {Esposito-Far{\`e}se},
  {Verbiest}, {Bailes}, {Jacoby}, {Kramer}, {Stairs}, {Antoniadis}, \&
  {Janssen}}]{fwe+12}
{Freire} P.~C.~C. {et~al.}, 2012, MNRAS, 423, 3328

\bibitem[{Hewish {et~al}\mbox{.}(1968)Hewish, Bell, Pilkington, Scott, \&
  Collins}]{hbp+68}
Hewish A., Bell S.~J., Pilkington J. D.~H., Scott P.~F., Collins R.~A., 1968,
  Nature, 217, 709

\bibitem[{Hills(1983)}]{hil83}
Hills J.~G., 1983, ApJ, 267, 322

\bibitem[{{Hobbs} {et~al}\mbox{.}(2004{\natexlab{a}}){Hobbs}, {Faulkner},
  {Stairs}, {Camilo}, {Manchester}, {Lyne}, {Kramer}, {D'Amico}, {Kaspi},
  {Possenti}, {McLaughlin}, {Lorimer}, {Burgay}, {Joshi}, \&
  {Crawford}}]{hfs+04}
{Hobbs} G. {et~al.}, 2004{\natexlab{a}}, MNRAS, 352, 1439

\bibitem[{{Hobbs} {et~al}\mbox{.}(2009){Hobbs}, {Hollow}, {Champion}, {Khoo},
  {Yardley}, {Carr}, {Keith}, {Jenet}, {Amy}, {Burgay}, {Burke-Spolaor},
  {Chapman}, {Danaia}, {Homewood}, {Kovacevic}, {Mao}, {McKinnon}, {Mulcahy},
  {Oslowski}, \& {van Straten}}]{hlc+09}
{Hobbs} G. {et~al.}, 2009, PASA, 26, 468

\bibitem[{{Hobbs} {et~al}\mbox{.}(2004{\natexlab{b}}){Hobbs}, {Manchester},
  {Teoh}, \& {Hobbs}}]{hmth04}
{Hobbs} G., {Manchester} R., {Teoh} A., {Hobbs} M., 2004{\natexlab{b}}, in
  Young Neutron Stars and Their Environments, {IAU} Symposium 218, Camilo F.,
  Gaensler B.~M., eds., Astronomical Society of the Pacific, San Francisco, pp.
  139--140

\bibitem[{{Jacoby} {et~al}\mbox{.}(2009){Jacoby}, {Bailes}, {Ord}, {Edwards},
  \& {Kulkarni}}]{jbo+09}
{Jacoby} B.~A., {Bailes} M., {Ord} S.~M., {Edwards} R.~T., {Kulkarni} S.~R.,
  2009, ApJ, 699, 2009

\bibitem[{{Keith} {et~al}\mbox{.}(2009){Keith}, {Eatough}, {Lyne}, {Kramer},
  {Possenti}, {Camilo}, \& {Manchester}}]{kel+09}
{Keith} M.~J., {Eatough} R.~P., {Lyne} A.~G., {Kramer} M., {Possenti} A.,
  {Camilo} F., {Manchester} R.~N., 2009, MNRAS, 395, 837

\bibitem[{{Keith} {et~al}\mbox{.}(2010){Keith}, {Jameson}, {van Straten},
  {Bailes}, {Johnston}, {Kramer}, {Possenti}, {Bates}, {Bhat}, {Burgay},
  {Burke-Spolaor}, {D'Amico}, {Levin}, {McMahon}, {Milia}, \&
  {Stappers}}]{kjv+10}
{Keith} M.~J. {et~al.}, 2010, MNRAS, 409, 619

\bibitem[{{Knispel} {et~al}\mbox{.}(2013){Knispel}, {Eatough}, {Kim}, {Keane},
  {Allen}, {Anderson}, {Aulbert}, {Bock}, {Crawford}, {Eggenstein}, {Fehrmann},
  {Hammer}, {Kramer}, {Lyne}, {Machenschalk}, {Miller}, {Papa}, {Rastawicki},
  {Sarkissian}, {Siemens}, \& {Stappers}}]{kek+13}
{Knispel} B. {et~al.}, 2013, ArXiv e-prints

\bibitem[{{Knispel} {et~al}\mbox{.}(2011){Knispel}, {Lazarus}, {Allen},
  {Anderson}, {Aulbert}, {Bhat}, {Bock}, {Bogdanov}, {Brazier}, {Camilo},
  {Chatterjee}, {Cordes}, {Crawford}, {Deneva}, {Desvignes}, {Fehrmann},
  {Freire}, {Hammer}, {Hessels}, {Jenet}, {Kaspi}, {Kramer}, {van Leeuwen},
  {Lorimer}, {Lyne}, {Machenschalk}, {McLaughlin}, {Messenger}, {Nice}, {Papa},
  {Pletsch}, {Prix}, {Ransom}, {Siemens}, {Stairs}, {Stappers}, {Stovall}, \&
  {Venkataraman}}]{kla+11}
{Knispel} B. {et~al.}, 2011, ApJ, 732, L1

\bibitem[{{Kramer} {et~al}\mbox{.}(2003){Kramer}, {Bell}, {Manchester}, {Lyne},
  {Camilo}, {Stairs}, {D'Amico}, {Kaspi}, {Hobbs}, {Morris}, {Crawford},
  {Possenti}, {Joshi}, {McLaughlin}, {Lorimer}, \& {Faulkner}}]{kbm+03}
{Kramer} M. {et~al.}, 2003, MNRAS, 342, 1299

\bibitem[{Kramer {et~al}\mbox{.}(2006)Kramer, Stairs, Manchester, McLaughlin,
  Lyne, Ferdman, Burgay, Lorimer, Possenti, D'Amico, Sarkissian, Hobbs,
  Reynolds, Freire, \& Camilo}]{ksm+06}
Kramer M. {et~al.}, 2006, Science, 314, 97

\bibitem[{Kulkarni \& Narayan(1988)}]{kn88}
Kulkarni S.~R., Narayan R., 1988, ApJ, 335, 755

\bibitem[{{Lazaridis} {et~al}\mbox{.}(2009){Lazaridis}, {Wex}, {Jessner},
  {Kramer}, {Stappers}, {Janssen}, {Desvignes}, {Purver}, {Cognard},
  {Theureau}, {Lyne}, {Jordan}, \& {Zensus}}]{lwj+09}
{Lazaridis} K. {et~al.}, 2009, MNRAS, 400, 805

\bibitem[{{Lee} {et~al}\mbox{.}(2013){Lee}, {Stovall}, {Jenet}, {Martinez},
  {Dartez}, {Mata}, {Lunsford}, {Cohen}, {Biwer}, {Rohr}, {Flanigan}, {Walker},
  {Banaszak}, {Allen}, {Barr}, {Bhat}, {Bogdanov}, {Brazier}, {Camilo},
  {Champion}, {Chatterjee}, {Cordes}, {Crawford}, {Deneva}, {Desvignes},
  {Ferdman}, {Freire}, {Hessels}, {Karuppusamy}, {Kaspi}, {Knispel}, {Kramer},
  {Lazarus}, {Lynch}, {Lyne}, {McLaughlin}, {Ransom}, {Scholz}, {Siemens},
  {Spitler}, {Stairs}, {Tan}, {van Leeuwen}, \& {Zhu}}]{lsj+13}
{Lee} K.~J. {et~al.}, 2013, ArXiv e-prints

\bibitem[{{Levin} {et~al}\mbox{.}(2013){Levin}, {Bailes}, {Barsdell}, {Bates},
  {Bhat}, {Burgay}, {Burke-Spolaor}, {Champion}, {Coster}, {D'Amico},
  {Jameson}, {Johnston}, {Keith}, {Kramer}, {Milia}, {Ng}, {Possenti},
  {Stappers}, {Thornton}, \& {van Straten}}]{lbb+13}
{Levin} L. {et~al.}, 2013, MNRAS

\bibitem[{Lorimer(1995)}]{lor95}
Lorimer D.~R., 1995, MNRAS, 274, 300

\bibitem[{Lorimer(2008)}]{lor08}
Lorimer D.~R., 2008, Living Reviews in Relativity, 11

\bibitem[{{Lorimer}(2013)}]{lor13}
{Lorimer} D.~R., 2013, in IAU Symposium, Vol. 291, IAU Symposium, pp. 237--242

\bibitem[{{Lorimer}, {Camilo} \& {McLaughlin}(2013){Lorimer}, {Camilo}, \&
  {McLaughlin}}]{lcm13}
{Lorimer} D.~R., {Camilo} F., {McLaughlin} M.~A., 2013, MNRAS

\bibitem[{{Lorimer} {et~al}\mbox{.}(2006){Lorimer}, {Faulkner}, {Lyne},
  {Manchester}, {Kramer}, {McLaughlin}, {Hobbs}, {Possenti}, {Stairs},
  {Camilo}, {Burgay}, {D'Amico}, {Corongiu}, \& {Crawford}}]{lfl+06}
{Lorimer} D.~R. {et~al.}, 2006, MNRAS, 372, 777

\bibitem[{Lorimer \& Kramer(2005)}]{lk05}
Lorimer D.~R., Kramer M., 2005, {Handbook of Pulsar Astronomy}. Cambridge
  University Press

\bibitem[{{Lynch} {et~al}\mbox{.}(2013){Lynch}, {Boyles}, {Ransom}, {Stairs},
  {Lorimer}, {McLaughlin}, {Hessels}, {Kaspi}, {Kondratiev}, {Archibald},
  {Berndsen}, {Cardoso}, {Cherry}, {Epstein}, {Karako-Argaman}, {McPhee},
  {Pennucci}, {Roberts}, {Stovall}, \& {van Leeuwen}}]{lbr+13}
{Lynch} R.~S. {et~al.}, 2013, ApJ, 763, 81

\bibitem[{Lyne {et~al}\mbox{.}(2004)Lyne, Burgay, Kramer, Possenti, Manchester,
  Camilo, McLaughlin, Lorimer, D'Amico, Joshi, Reynolds, \& Freire}]{lbk+04}
Lyne A.~G. {et~al.}, 2004, Science, 303, 1153

\bibitem[{Lyne {et~al}\mbox{.}(1998)Lyne, Manchester, Lorimer, Bailes, D'Amico,
  Tauris, Johnston, Bell, \& Nicastro}]{lml+98}
Lyne A.~G. {et~al.}, 1998, MNRAS, 295, 743

\bibitem[{Manchester {et~al}\mbox{.}(2001)Manchester, Lyne, Camilo, Bell,
  Kaspi, D'Amico, McKay, Crawford, Stairs, Possenti, Morris, \&
  Sheppard}]{mlc+01}
Manchester R.~N. {et~al.}, 2001, MNRAS, 328, 17

\bibitem[{Manchester {et~al}\mbox{.}(1996)Manchester, Lyne, D'Amico, Bailes,
  Johnston, Lorimer, Harrison, Nicastro, \& Bell}]{mld+96}
Manchester R.~N. {et~al.}, 1996, MNRAS, 279, 1235

\bibitem[{Manchester {et~al}\mbox{.}(1991)Manchester, Lyne, Robinson, D'Amico,
  Bailes, \& Lim}]{mlr+91}
Manchester R.~N., Lyne A.~G., Robinson C., D'Amico N., Bailes M., Lim J., 1991,
  Nature, 352, 219

\bibitem[{{Mickaliger} {et~al}\mbox{.}(2012){Mickaliger}, {Lorimer}, {Boyles},
  {McLaughlin}, {Collins}, {Hough}, {Tehrani}, {Tenney}, {Liska}, \&
  {Swiggum}}]{mlb+12}
{Mickaliger} M.~B. {et~al.}, 2012, ApJ, 759, 127

\bibitem[{{Morris} {et~al}\mbox{.}(2002){Morris}, {Hobbs}, {Lyne}, {Stairs},
  {Camilo}, {Manchester}, {Possenti}, {Bell}, {Kaspi}, {Amico}, {McKay},
  {Crawford}, \& {Kramer}}]{mhl+02}
{Morris} D.~J. {et~al.}, 2002, MNRAS, 335, 275

\bibitem[{{Os{\l}owski} {et~al}\mbox{.}(2011){Os{\l}owski}, {van Straten},
  {Hobbs}, {Bailes}, \& {Demorest}}]{ovh+11}
{Os{\l}owski} S., {van Straten} W., {Hobbs} G.~B., {Bailes} M., {Demorest} P.,
  2011, MNRAS, 418, 1258

\bibitem[{{Papitto} {et~al}\mbox{.}(2013){Papitto}, {Ferrigno}, {Bozzo}, {Rea},
  {Pavan}, {Campana}, {Romano}, {Burderi}, {Di Salvo}, {Riggio}, {Torres},
  {Falanga}, {Hessels}, {Burgay}, {Sarkissian}, {Wieringa}, {Filipovi{\'c}}, \&
  {Wong}}]{pfb+13}
{Papitto} A. {et~al.}, 2013, ArXiv e-prints

\bibitem[{Phinney \& Sigurdsson(1991)}]{ps91}
Phinney E.~S., Sigurdsson S., 1991, Nature, 349, 220

\bibitem[{Portegies~Zwart \& Yungelson(1998)}]{py98}
Portegies~Zwart S.~F., Yungelson L.~R., 1998, A\&A, 332, 173

\bibitem[{{Ransom} {et~al}\mbox{.}(2001){Ransom}, {Greenhill}, {Herrnstein},
  {Manchester}, {Camilo}, {Eikenberry}, \& {Lyne}}]{rgh+01}
{Ransom} S.~M., {Greenhill} L.~J., {Herrnstein} J.~R., {Manchester} R.~N.,
  {Camilo} F., {Eikenberry} S.~S., {Lyne} A.~G., 2001, ApJ, 546, L25

\bibitem[{{Ransom} {et~al}\mbox{.}(2005){Ransom}, {Hessels}, {Stairs},
  {Freire}, {Camilo}, {Kaspi}, \& {Kaplan}}]{rhs+05}
{Ransom} S.~M., {Hessels} J.~W.~T., {Stairs} I.~H., {Freire} P.~C.~C., {Camilo}
  F., {Kaspi} V.~M., {Kaplan} D.~L., 2005, Science, 307, 892

\bibitem[{{Ray} {et~al}\mbox{.}(2012){Ray}, {Abdo}, {Parent}, {Bhattacharya},
  {Bhattacharyya}, {Camilo}, {Cognard}, {Theureau}, {Ferrara}, {Harding},
  {Thompson}, {Freire}, {Guillemot}, {Gupta}, {Roy}, {Hessels}, {Johnston},
  {Keith}, {Shannon}, {Kerr}, {Michelson}, {Romani}, {Kramer}, {McLaughlin},
  {Ransom}, {Roberts}, {Saz Parkinson}, {Ziegler}, {Smith}, {Stappers},
  {Weltevrede}, \& {Wood}}]{rap+12}
{Ray} P.~S. {et~al.}, 2012, ArXiv e-prints

\bibitem[{{Roberts}(2011)}]{rob11}
{Roberts} M.~S.~E., 2011, in American Institute of Physics Conference Series,
  Vol. 1357, American Institute of Physics Conference Series, {Burgay} M.,
  {D'Amico} N., {Esposito} P., {Pellizzoni} A., {Possenti} A., eds., pp.
  127--130

\bibitem[{{Rosen} {et~al}\mbox{.}(2013){Rosen}, {Swiggum}, {McLaughlin},
  {Lorimer}, {Yun}, {Heatherly}, {Boyles}, {Lynch}, {Kondratiev}, {Scoles},
  {Ransom}, {Moniot}, {Cottrill}, {Weaver}, {Snider}, {Thompson}, {Raycraft},
  {Dudenhoefer}, {Allphin}, {Thorley}, {Meadows}, {Marchiny}, {Liska},
  {O'Dwyer}, {Butler}, {Bloxton}, {Mabry}, {Abate}, {Boothe}, {Pritt},
  {Alberth}, {Green}, {Crowley}, {Agee}, {Nagley}, {Sargent}, {Hinson},
  {Smith}, {McNeely}, {Quigley}, {Pennington}, {Chen}, {Maynard}, {Loope},
  {Bielski}, {McGough}, {Gural}, {Colvin}, {Tso}, {Ewen}, {Zhang},
  {Ciccarella}, {Bukowski}, {Novotny}, {Gore}, {Sarver}, {Johnson},
  {Cunningham}, {Collins}, {Gardner}, {Monteleone}, {Hall}, {Schweinhagen},
  {Ayers}, {Jay}, {Uosseph}, {Dunkum}, {Pal}, {Dydiw}, {Sterling}, \&
  {Phan}}]{rsm+13}
{Rosen} R. {et~al.}, 2013, ApJ, 768, 85

\bibitem[{Seymour, Lorimer \& Ridley(2013)Seymour, Lorimer, \& Ridley}]{slr13}
Seymour A., Lorimer D.~R., Ridley J.~P., 2013, MNRAS, submitted

\bibitem[{Sigurdsson \& Phinney(1990)}]{sp90}
Sigurdsson S., Phinney E.~S., 1990, BAAS, 22, 1341

\bibitem[{{Stappers} {et~al}\mbox{.}(2011){Stappers}, {Hessels}, {Alexov},
  {Anderson}, {Coenen}, {Hassall}, {Karastergiou}, {Kondratiev}, {Kramer}, {van
  Leeuwen}, {Mol}, {Noutsos}, {Romein}, {Weltevrede}, {Fender}, {Wijers},
  {B{\"a}hren}, {Bell}, {Broderick}, {Daw}, {Dhillon}, {Eisl{\"o}ffel},
  {Falcke}, {Griessmeier}, {Law}, {Markoff}, {Miller-Jones}, {Scheers},
  {Spreeuw}, {Swinbank}, {Ter Veen}, {Wise}, {Wucknitz}, {Zarka}, {Anderson},
  {Asgekar}, {Avruch}, {Beck}, {Bennema}, {Bentum}, {Best}, {Bregman},
  {Brentjens}, {van de Brink}, {Broekema}, {Brouw}, {Br{\"u}ggen}, {de Bruyn},
  {Butcher}, {Ciardi}, {Conway}, {Dettmar}, {van Duin}, {van Enst}, {Garrett},
  {Gerbers}, {Grit}, {Gunst}, {van Haarlem}, {Hamaker}, {Heald}, {Hoeft},
  {Holties}, {Horneffer}, {Koopmans}, {Kuper}, {Loose}, {Maat},
  {McKay-Bukowski}, {McKean}, {Miley}, {Morganti}, {Nijboer}, {Noordam},
  {Norden}, {Olofsson}, {Pandey-Pommier}, {Polatidis}, {Reich},
  {R{\"o}ttgering}, {Schoenmakers}, {Sluman}, {Smirnov}, {Steinmetz}, {Sterks},
  {Tagger}, {Tang}, {Vermeulen}, {Vermaas}, {Vogt}, {de Vos}, {Wijnholds},
  {Yatawatta}, \& {Zensus}}]{sha+11}
{Stappers} B.~W. {et~al.}, 2011, A\&A, 530, A80

\bibitem[{{Story}, {Gonthier} \& {Harding}(2007){Story}, {Gonthier}, \&
  {Harding}}]{sgh07}
{Story} S.~A., {Gonthier} P.~L., {Harding} A.~K., 2007, ApJ, 671, 713

\bibitem[{Tauris \& Bailes(1996)}]{tb96}
Tauris T.~M., Bailes M., 1996, A\&A, 315

\bibitem[{{Taylor} {et~al}\mbox{.}(2012){Taylor}, {Ellingson}, {Kassim},
  {Craig}, {Dowell}, {Wolfe}, {Hartman}, {Bernardi}, {Clarke}, {Cohen},
  {Dalal}, {Erickson}, {Hicks}, {Greenhill}, {Jacoby}, {Lane}, {Lazio},
  {Mitchell}, {Navarro}, {Ord}, {Pihlstr{\"o}m}, {Polisensky}, {Ray},
  {Rickard}, {Schinzel}, {Schmitt}, {Sigman}, {Soriano}, {Stewart}, {Stovall},
  {Tremblay}, {Wang}, {Weiler}, {White}, \& {Wood}}]{tek+12}
{Taylor} G.~B. {et~al.}, 2012, Journal of Astronomical Instrumentation, 1,
  50004

\bibitem[{Taylor \& Huguenin(1969)}]{th69}
Taylor J.~H., Huguenin G.~R., 1969, Nature, 221, 816

\bibitem[{Wolszczan(1991)}]{wol91a}
Wolszczan A., 1991, Nature, 350, 688

\end{thebibliography}
\end{document}